\documentclass[prb,preprint]{revtex4-1} 

\usepackage{amsmath} 
\usepackage{amsfonts} 
\usepackage{amssymb}
\usepackage{graphicx} 

\newcommand{\angbod}{\Omega_b}
\newcommand{\accbod}{\dot{\Omega}_b}
\newcommand{\anglab}{\Omega_l}
\newcommand{\tr}{
\mathop{\mbox{\rm Tr}}
}

\begin{document}

\title{
Understanding rigid body motion in arbitrary dimensions
}
\author{Francois Leyvraz}
\altaffiliation[Also at ]{Centro Internacional de Ciencias, Cuernavaca, M\'exico.} 
\affiliation{Instituto de Ciencias F\'\i sicas, UNAM, Cuernavaca, M\'exico}
\email{leyvraz@fis.unam.mx} 

\date{\today}

\begin{abstract}
Why would anyone wish to generalize the already unappetizing subject of 
rigid body motion to an arbitrary number of dimensions? At first sight, 
the subject seems to be both repellent and superfluous. The author will
try to argue that 
an approach involving no specifically three-dimensional constructs is actually easier
to grasp than the traditional one
and might thus be generally useful to understand rigid body motion
both in three dimensions and in the general case.
Specific differences between the viewpoint suggested here and the usual one 
include the following: here angular velocities are systematically
treated as antisymmetric matrices, a symmetric
tensor $I$ quite different from the moment of inertia tensor plays a central role, whereas the 
latter is shown to be a far more complex object, namely a tensor of rank four.
A straightforward way to define it is given. The Euler equation is derived and 
the use of Noether's theorem to obtain conserved quantities is illustrated. 
Finally the equations of motion for a heavy top as well as
for two bodies linked by a spherical joint are 
derived to display the simplicity and the power of the method. 
\end{abstract}

\maketitle

\section{Introduction}
Rigid body motion is one of the jewels of classical mechanics: it gives a straightforward
description of a system which behaves in an unexpected and often counterintuitive manner.
Its practical importance is also greater than is sometimes suspected: it merits a 
thorough treatment by Laplace \cite{laplace}, for example, due to its importance in 
the description of the Earth's motion. Its applications, however, are nearly numberless:
gyroscopes, robots, computer animation, toys such as the eternally
fascinating top and many more. 

The classical presentation of the subject is remarkably beautiful and the 
accounts of it  given in
various textbooks on classical mechanics, such as\cite{goldstein, landau, jose, arnold}, 
are all rather similar, which surely indicates that this topic has
achieved a nearly perfect form: it all starts with kinematics, the definition of
angular velocity as an axial vector, and finally
 the relation between the axial vector of angular momentum
and the axial vector of angular velocity through the moment of inertia tensor,
from which an equation of motion such as the Euler equation is then derived. 

It is an essential feature of the classical treatment that it takes place entirely in three dimensions: 
the vector product and the concept of axial vector always play a central role. The moment of inertia tensor
also plays a vital role, and, as we shall see, it cannot be defined in the usual 
fashion for  rigid bodies in more than three dimensions. All these concepts derive
from the fact that a rotation 
in three dimensions is always characterized by an {\em axis\/} as well as an angle of
rotation around this axis. This is why rotational motion can be characterized---at
the infinitesimal level at least---by vectors, displaying the instantaneous axis of the
 motion. 

All of this is false in dimensions higher than three: a rotation in 4 dimensions, for example,
consists of two independent two-dimensional rotations, characterized by two angles
$\phi_1$ and $\phi_2$, taking place in two orthogonal two-dimensional spaces. Clearly, no 
concept of axis survives. Thus, what we are proposing here is a description of
rigid body motion that does not use axes. While this might look like an idle
exercise, it turns out that the algebra becomes considerably more transparent
if we only limit ourselves to concepts that can be extended without difficulty to
arbitrary dimensions. While the algebra becomes simpler, it must be admitted
that the geometric intuition becomes rather less clear. However, this author, 
at least, has never found the classical treatment of the general case
to be very clear geometrically. Ever since Lagrange \cite{lagmech}
proudly stated that his work contained no figures and relied solely on calculation, there 
has been a fruitful tension between those who emphasize the geometric aspects of mechanics
and those more inclined to algebra. This paper, as will become clear, is squarely within 
the tradition of Lagrange. It may thus be of interest for a certain class of readers.

At this stage, it should of course be emphasized that the results are not new. 
Certainly Euler's equation for the $n$-dimensional free rigid body has been stated 
and analysed before. The first to propose the problem of the generalization of
the Euler equation to $n$ dimensions was Cayley \cite{cay46}. This problem was then solved 
by Frahm \cite{frahm}. Independently, a similar extension was made by Weyl, 
as an ``exercise'' in the use of tensor calculus \cite{weyl}. Finally, 
a derivation in modern terms was given by Arnol'd \cite{arnold66}. 
In contradistinction to the three-dimensional case, these
higher dimensional Euler equations are not obviously integrable. This issue
has thus attracted substantial mathematical interest. Among the first
results proved in this respect was the integrability of the four-dimensional
free Euler equations, shown by Schottky\cite{schott}.  
Non-trivial integrals of motion for arbitrary $n$ were derived in 
 \cite{mish70, man76}. These lead to the result that that 
the $n$ dimensional Euler equation is integrable in arbitrary 
dimensions. These results, however, are beyond the scope of this paper. 
It should, however, be pointed out that the subject has generated a large
mathematical literature, of which a small, unsystematic selection may
get the interested reader started: various results concerning the free 
$n$ dimensional top are found in \cite{ratiu, fedorov, blochetal}. Concerning the
$n$-dimensional heavy top, the reader will find further work in \cite{heavy1, heavy2}. 
It should, however, be emphasized that the works just cited make considerable use of 
differential geometry and require some significant mathematical background to 
be understood.

The difference between these works and the present paper is 
the point of view adopted: I am attempting to show that such an approach
can be made quite elementary and shall use no advanced mathematics whatever in the following. 
Further, I hope to make clear that this approach can be used to provide 
a better understanding of ordinary three-dimensional rigid body motion. 
I have recently become aware of another such elementary treatment
of this subject\cite{sinclair1, sinclair2}, however I believe the present approach is 
still sufficiently different to stand on its own.

It must, of course, be emphasized that there exist many attempts to clarify 
rigid body motion in a way entirely diferent from that pursued here, namely
by improving our geometric understanding of the 
three-dimensional system, specifically of rotations in three-dimensional space. 
This work is associated with the names of Poinsot\cite{poins51},
Klein and Sommerfeld\cite{klein} and has been pursued by a large number of workers, of which space 
only allows to cite a few\cite{hill45,grub62,mott66,thur67,grub72,edw77,leub79,vol93,siv94,beatt96}.

In Section \ref{sec:newton} we shall first review rapidly the kinematics of rigid body motion
in the terms best adapted to the general $n$-dimensional framework we are interested in. 
We then display a general Lagrangian for the rigid body, without using
specific coordinates such as the Euler angles. 
We proceed to derive a quite general equation which might be called
Newton's equation for rigid body motion. In Section \ref{sec:euler} we obtain 
from the equation of motion a generalized
version of Euler's equations and define the $n$-dimensional version
of the moment of inertia tensor. In Section \ref{sec:noether} we apply Noether's theorem to 
obtain expressions for the angular momentum of a rigid body, as well as
the conservation laws that follow from the symmetry of the rigid body. 
In Section \ref{sec:art} we first analyse the two-dimensional case.
We then present two non-trivial examples of our formalism: first,
we derive the equations of motion for a heavy top, that is, a rigid body suspended 
at a point different from its center of mass and subjected to a constant force. Second
we consider the case of two rigid bodies 
linked at one point by a frictionless spherical joint. 
In Section \ref{sec:conclusion} we present some conclusions. 

\section{Newton's equations for the rigid body}
\label{sec:newton}
The goal of this Section is to arrive at an equation of motion
for a rigid body under the influence of 
an arbitrary potential, which is the equivalent of Newton's equation for
a particle: it is a set of equations of second order involving and determining
uniquely all those parameters which describe the orientation of the rigid body. We first
begin by some elementary definitions, then define a Lagrangian and finally derive the 
equation of motion.
\subsection{Kinematics: rotations as coordinates}
The first difficulty in rigid body motion is to characterize the 
orientation of our system. A rigid body consists of an arbitrary 
number of particles linked by the constraint that all interparticle distances 
remain constant. A naive description would therefore involve a possibly
quite large number of particle coordinates with a comparably large 
number of constraints. 

To bypass this difficulty, we define a reference body at rest, and describe the 
configuration of the moving body by applying to the reference body
a time-dependent rotation 
$R(t)$ followed by a time-dependent  translation $\vec X(t)$. Thus, if the body 
consists of $N$ bodies of masses $m_i$, with $1\leq i\leq N$ in the positions
$\vec x_i(t)$, then there exist fixed positions $\vec a_i$ as well
as a rotation $R(t)$ and a vector $\vec X(t)$ such that 
\begin{equation}
\vec x_i(t)=R(t)
\vec a_i+\vec X(t).
\label{eq:2.1}
\end{equation}
While this certainly appears intuitively clear, a rigorous proof is not obvious: the 
interested reader is referred to 
\cite{thur67,beatt96}. It follows that the kinetic energy is given by
\begin{equation}
T=\frac12\sum_{i=1}^N
m_i\left(\dot{\vec x}_i,\dot{\vec x}_i\right)=\frac12\sum_{i=1}^Nm_i\left(
\dot R(t)\vec a_i+\dot{\vec X}(t),\dot R(t)\vec a_i+\dot{\vec X}(t)
\right),
\label{eq:2.2}
\end{equation}
which can also trivially be reformulated in the case of a continuous mass distribution.
In the following, for simplicity's sake, we shall always assume either that $\vec X(t)=0$
or that the origin of the rotations is taken at the center of mass of the
body. We are thus assuming that the origin is always taken at the center of mass, except
occasionally, when one point of the rigid body is fixed
at some point, which we then take as the origin of rotations. This leads us 
to the two following expressions for the kinetic energy
\begin{subequations}
\label{eq:2.2.1}
\begin{eqnarray}
T&=&\frac12\sum_{i=1}^Nm_i\left(
\dot R(t)\vec a_i,\dot R(t)\vec a_i\right)
+
\frac M2\left(
\dot{\vec X},
\dot{\vec X}
\right)\label{eq:2.2.1a}\\
M&=&\sum_{i=1}^N m_i
\label{eq:2.2.1b}\\
T&=&
\frac12\sum_{i=1}^Nm_i\left(
\dot R(t)\vec a_i,\dot R(t)\vec a_i
\right)\label{eq:2.2.1c}
\end{eqnarray}
\end{subequations}

Finally, let us quickly state explicitly a few elementary properties 
of rotations which we shall need in the following. The definition of a rotation $R$
is that, for every $\vec x$ and $\vec y$, one has
\begin{equation}
\left(
R\vec x,R\vec y
\right)=(\vec x,\vec y),
\label{eq:2.3}
\end{equation}
which follows from the definition of a rotation
as a linear map that leaves distances---and hence 
angles---invariant. From equation (\ref{eq:2.3}) follows that,
for an arbitrary rotation $R$
\begin{equation}
R^TR={\mathbb I}
\label{eq:2.5}
\end{equation}
where $\mathbb I$ is the identity matrix and $R^T$ denotes the transpose matrix. 
Let us now consider an 
arbitrary time-dependent rotation $R(t)$. Differentiating 
equation (\ref{eq:2.5}) for $R(t)$ with respect to $t$, yields
\begin{equation}
R^T(t)\dot R(t)+\dot R^T(t)R(t)=0,
\label{eq:2.6}
\end{equation}
from which immediately follows that the matrices
\begin{subequations}
\label{eq:2.7}
\begin{eqnarray}
\angbod(t)&=&R^{-1}(t)\dot{R}(t)\label{eq:2.7a}\\
\anglab(t)&=&\dot{R}(t)R^{-1}(t)\label{eq:2.7b}
\end{eqnarray}
\end{subequations}
are both {\em antisymmetric}. 

These antisymmetric matrices have a very remarkable significance
when $R(t)$ represents the motion of a rigid body according to 
equation (\ref{eq:2.1}) and both
will play a crucial role in all of what follows. 
To understand their physical meaning, 
first note that $R(t)$ maps the reference body on the 
moving body. We may therefore say that $\dot R(t)$ maps the positions
$\vec a_i$ of the reference body to the velocity of the corresponding point 
$\vec{x}_i(t)$. The matrix $\angbod$ therefore maps the positions 
$\vec a_i$ to the velocity which the point $\vec a_i$ would have,
if the reference body moved similarly to the physical body.
Similarly, the matrix $\anglab$ maps the points $\vec x_i(t)$ to the 
velocity of $\vec x_i(t)$. The matrix $\angbod$ is thus called the
{\em angular velocity in the body frame}, whereas the matrix $\anglab$
is called the {\em angular velocity in the laboratory frame}.
\subsection{The rigid body Lagrangian}
We now derive a Lagrangian for a rigid body in a general potential.
The coordinate describing the orientation of the rigid body 
is the rotation $R(t)$. We assume the potential energy
to depend solely on $R$, so that we need only focus on the expression for the kinetic
energy. We have already written an expression for it, see Eq.~(\ref{eq:2.2.1}). We now rewrite 
it as follows:
\begin{equation}
T=\frac12\sum_{\alpha,\beta,\gamma=1}^n\sum_{i=1}^N
m_i \dot R_{\alpha\beta}a_{i,\beta} \dot R_{\alpha\gamma}a_{i,\gamma}.
\label{eq:3.1}
\end{equation}
At this stage, we may disregard the term describing center of mass motion in
(\ref{eq:2.2.1a}). 
Let us make here the following remarks concerning notation, which we shall
stick to throughout the paper: Latin indices refer to particles and run from 1 to $N$, whereas
Greek indices from the beginning of the alphabet
refer to coordinates in the $n$-dimensional space in which the motion takes place, 
and thus run from 1 to $n$. We shall always use the notation $a_{i,\alpha}$ to refer to
the component $\alpha$ of the vector $\vec a_i$. After these remarks, equation (\ref{eq:3.1})
should be a straightforward consequence of Eq.~(\ref{eq:2.2.1}). 

We now define the following basic object, namely the {\em tensor of  second moments}:
\begin{equation}
I_{\alpha\beta}=\sum_{i=1}^N m_ia_{i,\alpha}a_{i,\beta}
\label{eq:3.2}
\end{equation}
An important remark: this is quite different from the moment of inertia tensor in
the traditional approach.
Thus  $I_{11}$ is large when the body is 
extended in the direction of the $x_1$ axis. In contradistinction to this, for the $11$ component
of the tensor of inertia to be large, the body must be extended in the $23$-plane.  
Another obvious difference between  $I$ and the 
moment of inertia tensor is the fact that, for $n=2$, $I$ is a $2\times2$ matrix, 
not a number.

We can now use the tensor $I$, see Eq.~(\ref{eq:3.2}), to simplify (\ref{eq:3.1}) considerably:
\begin{equation}
T=\frac12\tr\left(
\dot RI\dot{R}^T
\right)
.
\label{eq:3.4}
\end{equation}
For a definition of the trace as well as some useful elementary properties, see Appendix \ref{sec:trace}. 
This, together with the potential term $V(R)$ defines the Lagrangian, up to an important issue: we
have not specified explicitly that the matrices
$R$ must be rotations. The easiest way to do this 
is by imposing a Lagrange multiplier $\Lambda$, which is an $n\times n$ matrix. 
The condition for $R$ to be a rotation can be stated via Eq.~(\ref{eq:2.6}), so that 
the final Lagrangian describing the rotational motion
is given by:
\begin{equation}
L(\dot R, R)=\frac12\tr
\left(
\dot RI\dot{R}^T\right)
+\tr
\left[
\Lambda\left(
R^T\dot R+\dot{R}^TR
\right)
\right]-V(R)
.
\label{eq:3.6}
\end{equation} 
A very important observation should be made at this stage: since the matrix 
$R^T\dot R+\dot{R}^TR$
is automatically symmetric, $\Lambda$ can without loss of generality be assumed 
symmetric also.

\subsection{Equations of motion}

We now proceed to derive the equations of motion from the Lagrangian (\ref{eq:3.6}).
Using the various tricks described in Appendix \ref{sec:trace}, we readily obtain:
\begin{subequations}
\begin{eqnarray}
\frac{\partial L(\dot R, R)}{\partial \dot R}&=&
\dot RI+2R\Lambda
\label{eq:3.7a}\\
\frac{\partial L(\dot R, R)}{\partial R}&=&2\dot R\Lambda-\frac{\partial V(R)}{\partial R}.
\end{eqnarray}
\label{eq:3.7}
\end{subequations}
Writing down the Euler--Lagrange equations thus yields
\begin{equation}
\ddot RI+2R\dot{\Lambda}=-\frac{\partial V(R)}{\partial R}.
\label{eq:3.8}
\end{equation}
It now remains to eliminate the $\Lambda$. This is done via the observation made immediately after
Eq.~(\ref{eq:3.6}), that $\Lambda=\Lambda^T$. Multiplying (\ref{eq:3.8}) on the left by $R^{-1}$
and rearranging, one finds
\begin{equation}
R^{-1}\ddot RI+R^{-1}\left(
\frac{\partial V(R)}{\partial R}
\right)
=-2\dot{\Lambda}.
\label{eq:3.9}
\end{equation}
From this follows that the left-hand side of (\ref{eq:3.9}) is symmetric, that is,
using the fact that $R^{-1}=R^T$ and antisymmetrizing:
\begin{equation}
R^{T}\ddot RI-I\ddot{R}^TR =
\left(\frac{\partial V(R)}{\partial R}\right)^TR
-R^T\left(
\frac{\partial V(R)}{\partial R}
\right)
.
\label{eq:3.11}
\end{equation}
At this point we should pause to ask the meaning of such quantities as $\partial V/\partial R$
and in particular the right-hand side of (\ref{eq:3.11}). Since 
$\tr(AB^T)$ defines a scalar product among matrices (see (\ref{eq:a7})), we can 
define  $\partial V/\partial R$ by the relation
\begin{equation}
 V(R+\delta R)-V(R)\simeq\tr\left[
\left(
\frac{\partial V(R)}{\partial R}
\right)
\left(
 \delta R
 \right)^T
 \right]
 , 
\label{eq:3.11a}
\end{equation}
where $\delta R$ is a small matrix with the property that $R+\delta R$ is still a rotation. It is thus
in first order of the form $R\delta A$, where $\delta A$ is an infinitesimal antisymmetric matrix. 
We may thus rewrite (\ref{eq:3.11a}) as 
\begin{eqnarray}
V(R+R\delta A)-V(R)&\simeq&\tr\left[
\left(
R^T\frac{\partial V(R)}{\partial R}
\right)
\left(
\delta A
\right)^T
\right]\nonumber\\
&=&\frac12
\tr\left[
\left(
R^T\frac{\partial V(R)}{\partial R}
-
\frac{\partial V(R)}{\partial R}^TR
\right)
\left(
\delta A
\right)^T
\right],
\label{eq:3.11b}
\end{eqnarray}
where in the final step we use the antisymmetry of $\delta A$. The right-hand side of (\ref{eq:3.11})
is thus---up to a sign and a factor of 2---the change in energy caused by an infinitesimal rotation in the body frame. 

We now claim that equation (\ref{eq:3.11}) can justly be viewed as ``Newton's equations for a rigid body''. They are
a set of second-order differential equations, the number of which is
exactly sufficient to
describe the dynamics of the rotation $R(t)$. Indeed, equation (\ref{eq:3.11}) states that 
two antisymmetric matrices are equal. 
The number of independent equations in equation (\ref{eq:3.11}) is therefore $n(n-1)/2$. That rotations are
described by the same number of independent parameters follows, for example, from the fact that 
rotations near the identity $\mathbb I$ are in first order equal to ${\mathbb I}+A$, where $A$ is an arbitrary 
antisymmetric matrix. For applications, we might now describe $R$ by our favourite parametrization---whether
Euler angles, quaternions or any other---and obtain equations of motion for these
 without further ado\cite{foot}.
%

\section{The Euler equation}
\label{sec:euler}
We now rewrite (\ref{eq:3.11}) as two equations of first order. As the first, we take the definition
of $\angbod$ given by (\ref{eq:2.7a}). We then use (\ref{eq:2.7a}) to express $\ddot{R}$: 
\begin{subequations}
\label{eq:4.1}
\begin{eqnarray}
\dot R&=&R\angbod\label{eq:4.1a}\\
\ddot R&=&\dot R\angbod+R\accbod=R\left(
\angbod^2+\accbod
\right)\label{eq:4.1b}
\end{eqnarray}
\end{subequations}
Putting Eq.~(\ref{eq:4.1b}) in the equation of motion (\ref{eq:3.11}), one obtains
\begin{equation}
I\accbod+\accbod I+\left(
\angbod^2I-I\angbod^2
\right)=\left(\frac{\partial V(R)}{\partial R}\right)^TR
-R^T\left(
\frac{\partial V(R)}{\partial R}
\right)
.
\label{eq:4.3}
\end{equation}
Here we have used the fact that $\accbod$ is antisymmetric and that $\angbod^2$
is symmetric.

If we define the commutator and the anticommutator of 2 matrices $A$ and $B$
\begin{subequations}
\begin{eqnarray}
\left[
A,B
\right]&=&AB-BA\label{eq:4.3.1a}\\
\left\{
A,B
\right\}
&=&AB+BA
,
\label{eq:4.3.1b}
\end{eqnarray}
\label{eq:4.3.1}
\end{subequations}
we can write Eq.~(\ref{eq:4.3}) as follows
\begin{equation}
\left\{
I,\accbod
\right\}+\left[
\angbod^2, I
\right]=\left(\frac{\partial V(R)}{\partial R}\right)^TR
-R^T\left(
\frac{\partial V(R)}{\partial R}
\right)
.
\label{eq:4.3.2}
\end{equation}
It is readily verified that the various commutators and anticommutators involved are all antisymmetric
as is the right-hand side of (\ref{eq:4.3.2}). 
Eq.~(\ref{eq:4.3.2}) is, as we shall see, the Euler equation with a torque term. We 
proceed to show that it can be written in a form reminiscent of the usual one. 

We thus look for an analogue of the moment of inertia
tensor for Eq.~(\ref{eq:4.3}). Let us define the {\em superoperator\/}
$\hat{\Theta}_I$
which maps every antisymmetric matrix $A$ to another such in the following manner:
\begin{equation}
\hat{\Theta}_I(A)=\left\{
I, A
\right\}=
IA+AI
\label{eq:4.4}
\end{equation}
$\hat{\Theta}_I$ is therefore an operator defined on the space of all $n\times n$
antisymmetric matrices, that is, on a space of dimension $n(n-1)/2$. It is hence a tensor of
fourth rank on the original space. Using 
this definition, one finds that (\ref{eq:4.3}) can be rewritten as
\begin{equation}
\hat{\Theta}_I(\accbod)+\left[
\angbod,\hat{\Theta}_I(\angbod)
\right]=
\left(\frac{\partial V(R)}{\partial R}\right)^TR
-R^T\left(
\frac{\partial V(R)}{\partial R}
\right)
.
\label{eq:4.5}
\end{equation}
which now looks quite similar to the usual form of the Euler equation
where the right-hand side is found to correspond to 
the torque---expressed in the body frame---exerted on the body by $V(R)$. 
That Eq.~(\ref{eq:4.5})
does in fact reduce to the ordinary Euler equation 
in three dimensions for the case in which $V(R)=0$ is shown in Appendix \ref{sec:euler3d}. 

Note that (\ref{eq:4.3.2}) is not in any way more simple than the original
Newton equation (\ref{eq:3.11}). But in the particular case that $V=0$, 
a simplification arises: the equation (\ref{eq:4.3.2}) becomes 
closed in $\angbod$, that is, we may, by solving (\ref{eq:4.3.2}), obtain $\angbod(t)$
from the initial angular velocities $\angbod(0)$. This is the Euler equation for free rigid body 
motion.

This equation yields the same kind of information as the usual Euler equation: 
for example, we see that there are permanent rotations, that is $\accbod=0$, if and only if
$I$ commutes with $\angbod^2$. As follows from elementary linear algebra, 
all the eigenspaces of $\angbod^2$ are two-dimensional, except possibly
for a one dimensional null zero eigenspace.  A rotation is therefore only permanent 
when all these eigenspaces---including the null space if it exists---are so chosen 
that the eigenvectors of $I$ lie in them. 
Such a statement is, of course, well-known to hold in three dimensions, 
though it is usually formulated somewhat differently.

Finally, we point out that solving the Euler equations (\ref{eq:4.5}) for $V=0$ 
does not mean that the motion 
of the system is known: to this end one needs to solve additionally the first order equation
\begin{equation}
\dot R(t)=R(t)\angbod(t).
\label{eq:4.6}
\end{equation}
Given $\angbod(t)$, this is a time-dependent system of ordinary linear 
differential equations, which cannot
be solved, save in exceptional cases, using the matrix exponential\cite{teschl}, since in general
\begin{equation}
\left[
\angbod(t),
\angbod(t^\prime)
\right]\neq0\qquad(t\neq t^\prime).
\end{equation}
However, it frequently happens that knowledge of $\angbod$ is sufficient. A geometric 
way of obtaining $R(t)$ for the three-dimensional case, is given by the celebrated Poinsot construction, which 
we shall not discuss further, though it can be generalized to 
arbitrary dimensions, see for example \cite{fedorov}. 
\section{Conserved quantities, Noether's theorem and angular momentum}
\label{sec:noether}
It is a standard theorem of mechanics that any symmetry of a Lagrangian
is associated to the presence of a conserved quantity associated to that symmetry. 
Let us briefly state the theorem, referring to\cite{goldstein, landau, jose, arnold} for a proof. 
A symmetry of a system described by the generalized coordinates $q_1,\ldots,q_f$
is defined as follows: let us consider a continuous transformation of the $q_k$ 
depending on  parameter
$\lambda$ and inducing a transformation on the velocity variables given by
\begin{subequations}
\label{eq:5.1}
\begin{eqnarray}
Q_k(q_1,\ldots,q_f;\lambda)&=&\Phi_k(q_1,\ldots,q_f;\lambda)\label{eq:5.1a}\\
\dot{Q}_k(q_1,\ldots,q_f;\lambda)&=&\sum_{l=1}^f\frac{\partial\Phi_k(q_1,\ldots,q_f;\lambda)}{\partial q_l}\dot{q}_l,
\label{eq:5.1b}
\end{eqnarray}
\end{subequations}
where $1\leq k\leq f$. 
Such a transformation is called a {\em symmetry\/} if it leaves the Lagrangian invariant, that 
is, if
\begin{equation}
L\left[
\left(
\dot{Q}_k(q_1,\ldots,q_f;\lambda)
\right)_{k=1}^f;
\left(
Q_k(q_1,\ldots,q_f;\lambda)
\right)_{k=1}^f
\right]
=
L
\left[
\left(\dot q_k\right)_{k=1}^f;
\left
(q_k
\right)_{k=1}^f
\right],
\label{eq:5.2}
\end{equation}
where $(q_k)_{k=1}^f$ denotes the list $(q_1,\ldots,q_f)$. 

In the presence of the symmetry defined by (\ref{eq:5.1}), it can be shown that the
following quantity is conserved:
\begin{equation}
s=\sum_{l=1}^f\left(
\frac{\partial L\left[
\dot{q}_1,\ldots,\dot{q}_f;
q_1,\ldots,q_f
\right]}{\partial\dot{q}_l}
\left.
\frac{\partial\Phi_l(q_1,\ldots,q_f;\lambda)}{\partial\lambda}
\right|_{\lambda=0}
\right).
\label{eq:5.3}
\end{equation}
In the following, we shall differentiate between scalar and matrix 
conserved quantities by denoting the former with lower case and the latter
with capitalized Latin letters. 

We apply this result to the Lagrangian (\ref{eq:3.6}). The following transformation is 
a symmetry if $V(R)=0$:
\begin{subequations}
\label{eq:5.4}
\begin{eqnarray}
\Phi(R;\lambda)&=&e^{\lambda\Omega_0}R\label{eq:5.4a}\\
\frac{\partial\Phi(R;\lambda)}{\partial\lambda}&=&\Omega_0e^{\lambda\Omega_0}\dot R,\label{eq:5.4b}
\end{eqnarray}
\end{subequations}
where $\Omega_0$ is a fixed antisymmetric matrix. Here we use the usual definition 
of the matrix exponential and remind the reader that the exponential of an antisymmetric 
matrix is always a rotation, as follows, say, by integrating 
any of the forms of (\ref{eq:2.7}). This can be done without problems, since all matrices
of the form $e^{\lambda\Omega_0}$ commute among each other. 

Since the $q_l$ in the above formulae
correspond to $R_{\alpha\beta}$, we see that the indices $l$ correspond to double indices
$\alpha\beta$ in our problem. One gets
\begin{equation}
s(\Omega_0)=\sum_{\alpha,\beta=1}^n\left(
\dot R I
\right)_{\alpha\beta}
\left(
\Omega_0R
\right)_{\alpha\beta}=\tr\left(
 \dot RI R^T\Omega_0^T
 \right).
\label{eq:5.5}
\end{equation}
We may now rewrite this as
\begin{equation}
s(\Omega_0)=-\tr
\left(
 \dot RI R^T\Omega_0
\right).
\label{eq:5.6}
\end{equation}
Since $\Omega_0$ is an arbitrary antisymmetric matrix, it follows that
the antisymmetric part of $\dot RI R^T$ is a (matrix) conserved quantity:
\begin{equation}
S=\dot RI R^T-RI\dot R^T.
\label{eq:5.7}
\end{equation}
Using the definitions of $\angbod$ and $\anglab$, we can give two interesting expressions for $S$:
\begin{subequations}
\label{eq:5.8}
\begin{eqnarray}
S&=&R\left(
\angbod I+I\angbod
\right)R^{-1}\label{eq:5.8a}\\
&=&
\anglab RIR^{-1}+RIR^{-1}\anglab
.
\label{eq:5.8b}
\end{eqnarray}
\end{subequations}
$S$ can thus finally be expressed in terms of $\angbod$ or $\anglab$ and an appropriate moment of
inertia:
\begin{subequations}
\label{eq:5.9}
\begin{eqnarray}
S&=&R\hat{\Theta}_I\left(
\angbod
\right)R^{-1}\label{eq:5.9a}\\
&=&\hat{\Theta}_{RIR^{-1}}
\left(
\anglab
\right)
\label{eq:5.9b}
\end{eqnarray}
\end{subequations}
Since $S$ is obtained from rotational invariance, it is identified as the
angular momentum of the system. We have shown that all its components are conserved in the free case.
If we have a potential $V(R)$, this will generally not be true any more. If 
$V(R)$ is symmetric under some group of rotations, however, say the rotations generated by a given 
$\Omega_0$, then $\tr\left( S\Omega_0\right)$, which might be called the ``$\Omega_0$ component''
of the angular momentum tensor $S$, is conserved. Note further that the tensor $S$
is an object that maps points belonging to the moving body to point belonging to the
moving body again, that is, it is an object defined in the laboratory frame. Of course, using the 
techniques described in Appendix \ref{sec:euler3d}, we obtain the usual
expression for the angular momentum in either frame for three-dimensional systems. 

From Eq.~(\ref{eq:5.9a}) we can rederive the Euler equation by writing out the conservation of angular momentum
in terms of $\angbod$. This is nothing else than the usual derivation presented in textbooks. 
For completeness' sake, we show it:
\begin{eqnarray}
0&=&\frac{dS}{dt}\nonumber\\
&=&\dot R\hat{\Theta}_I\left(
\angbod
\right)R^{-1}+R\hat{\Theta}_I\left(
\accbod
\right)R^{-1}+
R\hat{\Theta}_I\left(
\angbod
\right)\dot{R}^T\nonumber\\
&=&R\left(
\hat{\Theta}_I(\accbod)+\left[
\angbod,\hat{\Theta}_I(\angbod)
\right]
\right)
R^{-1},
\end{eqnarray}
from which Euler's equation (\ref{eq:4.5}) follows. 

We have defined the symmetry (\ref{eq:5.4}) by {\em premultiplying} $R$ by a constant
rotation. This is essential: if we instead attempt the transformation
\begin{subequations}
\label{eq:5.11}
\begin{eqnarray}
R(\lambda)&=&Re^{\lambda\Omega_0}\label{eq:5.11a}\\
\dot R(\lambda)&=&\dot R\,e^{\lambda\Omega_0},\label{eq:5.11b}
\end{eqnarray}
\end{subequations}
it is quite easy to check that this does not, in general, leave the Lagrangian (\ref{eq:3.6}) invariant.
It does so only if $I$ commutes with $\Omega_0$. If this happens, as is the 
case when the rigid body has some symmetry, then we can indeed derive a conservation
law from this symmetry. By an exactly analogous computation, we see that the
 corresponding conservation law is given by
\begin{equation}
\tilde s(\Omega_0)=-\tr\left( R^T\dot RI \Omega_0\right).
\label{eq:5.12}
\end{equation}
If we define $\tilde S$ by
\begin{equation}
\tilde S=R^T\dot RI -I\dot R^T R =\angbod I+I\angbod=\hat{\Theta}_I(\angbod),
\label{eq:5.13}
\end{equation}
then we see that the expression
\begin{equation}
\tr
\left(
\tilde S\Omega_0
\right)
\label{eq:5.14}
\end{equation}
is conserved whenever $\Omega_0$ commutes with $I$. This is, of course, the angular momentum defined
in the body frame, which is conserved for a symmetric free body, though not otherwise. 

\section{Two Examples}
\label{sec:art}

To show how the formalism described above works, let us first do a routine exercise:
we look at the case of a two-dimensional rigid body, for the description of which only 
one angle is needed. The formula for the rotation as a function of the angle is
\begin{equation}
R(\phi)=
\left(
\begin{array}{cc}
\cos\phi&\sin\phi\\
-\sin\phi&\cos\phi
\end{array}
\right)
\label{eq:6.0.0.1}
\end{equation}
and hence
\begin{equation}
\angbod=\left(
\begin{array}{cc}
0&\dot\phi\\
-\dot\phi&0
\end{array}
\right)=\anglab
\label{eq:6.0.0.2}
\end{equation}
Since we are dealing with antisymmetric $2\times2$ matrices, we may characterize them 
uniquely by their upper right matrix element. 
Note that $\angbod$ and $\anglab$ are still different conceptually, and cannot 
really be compared, as they act on different spaces. They are, however, numerically equal. 

In the Euler equation, the commutator term vanishes, so we are left with
the term $\hat{\Theta}_I(\angbod)$, which is simply the antisymmetric matrix corresponding 
to $I\dot\phi$. 
We may now introduce a {\em scalar potential\/} $V_s(\phi)$ defined as $V[R(\phi)]$, where $R(\phi)$ is defined 
via (\ref{eq:6.0.0.1}). From the discussion leading to (\ref{eq:3.11b}),
we see that the right-hand side of (\ref{eq:3.11}) is simply the antisymmetric matrix corresponding to
$-V_s^\prime(\phi)$. Euler's equation thus reduces to
\begin{equation}
I\ddot\phi=-V_s^\prime(\phi).
\label{eq:6.0.0.3}
\end{equation}
Deriving the corresponding expressions for energy and angular momentum is an easy exercise, 
best left to the reader. 

We now proceed to work out two less trivial examples in which the method described here leads
straightforwardly both to compact expressions for the equations of motion
as well as for the conservation laws. First, let us consider the heavy top, suspended
at an arbitrary point. We make no assumption of an axis or symmetry, nor any 
further assumption on the location of the center of mass on some principal axis. 

As coordinates we take only rotations $R$, assuming the top's point of suspension
to be the origin, so that no displacement $\vec X$ is needed. The tensor $I$ is thus computed 
from the suspension point. Defining $\vec a$ as the coordinate of the center of mass in the 
reference body and $\vec g$ to be the direction of the acceleration gravity, we have
for the potential
\begin{equation}
V(R)=-m\left(
\vec g, R\vec a
\right)=-m\tr
\left[
\left(
\vec a\otimes\vec g
\right)R
\right],
\label{eq:6.0.1}
\end{equation}
where $m$ is the mass of the body.
Obviously, both $\vec a$ and $\vec g$ can be chosen to be in the $z$-direction by an 
appropriate choice of orientation in both the reference bodies and the laboratory coordinate 
system. However we shall not do this, as the gain in clarity resulting from clearly separating 
body-fixed quantities such as $\vec a$ from laboratory quantities such as $\vec g$ outweighs any
advantage in having slightly shorter formulae. 

The equation of motion are thus:
\begin{subequations}
\label{eq:6.0.2}
\begin{eqnarray}
\hat{\Theta}_I(\accbod)+\left[
\angbod,\hat{\Theta}_I(\angbod)
\right]&=&m\left[
R^T\left(
\vec a\otimes\vec g
\right)^T-\left(
\vec a\otimes\vec g
\right)R
\right]
\label{eq:6.0.2a}\\
\dot R&=&R\angbod\label{6.0.2b}
\end{eqnarray}
\end{subequations}
If $\Omega_0$ is any antisymmetric matrix such that $\Omega_0\vec g=0$, then the quantity
\begin{equation}
s(\Omega_0)=-\tr
\left(
\dot RI R^T\Omega_0
\right)=-\frac12\tr
\left[
\left(
\dot RIR^T-RI\dot{R}^T
\right)\Omega_0=
-\frac12\tr\hat{\Theta}_{RIR^{-1}}\left(
\anglab
\right)\Omega_0
\right]
\label{eq:6.0.3}
\end{equation}
is conserved. If further the vector $\vec a$ is an eigenvector of $I$ and additionally 
an antisymmetric matrix $\Omega_1$ exists such that 
\begin{equation}
\Omega_1\vec a=0,\qquad \left[
\Omega_1, I
\right]=0,
\label{eq:6.0.4}
\end{equation}
then the quantity
\begin{equation}
\tilde s(\Omega_1)=-\tr\left( R^T\dot RI \Omega_1\right)=-\frac12\tr
\left(
\hat{\Theta}_I(\angbod)\Omega_1
\right)
\label{eq:6.0.5}
\end{equation}
is also conserved. This corresponds, of course, to the situation in the 
integrable Lagrange top in three dimensions. Indeed, in three dimensions, this gives 
two integrals of motion, which together with the energy yields enough integrals of
motion, which in the Hamiltonian formalism turn out to be in involution, to 
give the complete integrability of the system. Note that, in this respect, the higher 
dimensional systems are truly more complicated: first, there are several degrees of 
symmetry, depending on how many eigenvalues of $I$ are degenerate. Second, if
one counts the number of conserved quantities obtained in this way, one generally does
not have enough to guarantee integrability. In fact, even the integrability of the Euler
equation in $n$ dimensions is by no means obvious \cite{man76} and certainly does not follow
from rotational invariance alone. 

At first sight, this requires using some parametrization of the rotations, such as Euler angles.
Such an approach does indeed yield the usual equations, as described in \cite{goldstein,landau,jose}.
This is, however, not necessary, as the following easy observation shows: define
\begin{equation}
\vec\gamma(t)=mR(t)^T\vec g
\label{eq:new1}
\end{equation}
The equations then read
\begin{subequations}
\label{eq:new2}
\begin{eqnarray}
\hat{\Theta}_I(\accbod)+\left[
\angbod,\hat{\Theta}_I(\angbod)
\right]&=&\left(
\vec a\otimes\vec\gamma
\right)^T-
\vec a\otimes\vec\gamma
\label{eq:new2a}\\
\dot{\vec\gamma}&=&-\angbod
\vec\gamma
\label{eq:new2b}
\end{eqnarray}
\end{subequations}
In three dimensions this is readily rewritten as: 
\begin{subequations}
\label{eq:new3}
\begin{eqnarray}
\Theta\dot{\vec\omega}_b+
\vec\omega\wedge\Theta\vec{\omega}_b
&=&-
\vec a\wedge\vec\gamma
\label{eq:new3a}\\
\dot{\vec\gamma}&=&-\vec{\omega}_b\wedge
\vec\gamma
\label{eq:new3b}
\end{eqnarray}
\end{subequations}
Here, of course, $\vec{\omega}_b$ is the vector corresponding to $\angbod$ and $\Theta$ is the 
$3\times3$ matrix corresponding to the superoperator $\hat{\Theta}_I$
This form for the equations of 
the heavy top is not usually given in the classical textbooks \cite{goldstein, landau, jose, arnold}, 
but it appears, for example, in \cite{lagmech}, see in particular the Second Part, Section VI, Paragraph 3, number 52. 
It is also the form used by Sophie Kowalevski \cite{soph-kow} to derive the integrable case of the heavy top named after her. 

We  now give another non-trivial example, for which the method here discussed
is remarkably straightforward. Consider two rigid bodies free to move arbitrarily in space, except
for the constraint that they be freely linked at a joint. This is described as follows:
we denote by $\sigma$ an index taking the values 1 and 2 and referring to
the two rigid bodies. 
Each body is described by a given rotation $R_\sigma(t)$ with respect to its center of  mass
and a translation $\vec{X}_\sigma(t)$. Both bodies are characterized by 
a tensor of second moments $I_\sigma$. Finally, the fact that they are linked 
is expressed by the fact that there exist constant vectors $\vec{A}_\sigma$ 
in the two bodies of reference such that
\begin{equation}
R_1(t)\vec{A}_1+\vec{X}_1(t)=
R_2(t)\vec{A}_2+\vec{X}_2(t)
.
\label{eq:6.1}
\end{equation}
for all $t$. The Lagrangian is hence given by
\begin{eqnarray}
&&L(\dot R_1, \dot R_2,\dot{\vec{X}}_1,\dot{\vec{X}}_2;R_1,R_2,\vec{X}_1,\vec{X}_2)=\sum_{\sigma=1,2}
\left\{
\frac12\tr 
\left(
\dot{R}_\sigma I_\sigma \dot{R}_\sigma^T
\right)
+\tr
\left[
\Lambda_\sigma\left(
R_\sigma^T\dot R_\sigma+\dot{R}_\sigma^TR_\sigma
\right)
\right]
\right\}\nonumber\\
&&\qquad+\sum_{\sigma=1,2}\left[
\frac{M_\sigma}{2}\left(
\dot{\vec{X}}_\sigma,
\dot{\vec{X}}_\sigma
\right)
+\vec\lambda\cdot\left(
R_1(t)\vec{A}_1+\vec{X}_1(t)-
R_2(t)\vec{A}_2-\vec{X}_2(t)
\right)
\right]
.
\label{eq:6.2}
\end{eqnarray}
Here all the notation is familiar, except for the vector $\vec\lambda$, which is the Lagrange multiplier
imposing the constraint (\ref{eq:6.1}). 
From (\ref{eq:a9}) of Appendix \ref{sec:trace} we can express the scalar product in a more convenient way: 
\begin{equation}
\vec\lambda\cdot\left(
R\vec{A}_\sigma
\right)=\tr
\left[
\left(
\vec\lambda\otimes\vec{A}_\sigma
\right)
R^T_\sigma
\right]
,
\label{eq:6.2b}
\end{equation}
which allows to use the techniques given in Appendix \ref{sec:trace} for these terms as well. 

The Euler--Lagrange equations are obtained in just the same way as in Section \ref{sec:newton} for the
rotational part. The part involving translations requires no further comment:
\begin{subequations}
\label{eq:6.3}
\begin{eqnarray}
M_1\ddot{\vec{X}}_1&=&\vec\lambda\label{eq:6.3a}\\
M_2\ddot{\vec{X}}_2&=&-\vec\lambda\label{eq:6.3b}\\
R_1 ^{T}\ddot{R}_1  I_1 -I_1 \ddot{R}_1 ^TR_1 &=&R^T_1 
\left(\vec\lambda\otimes\vec{A}_1 
\right)-
\left(
\vec\lambda\otimes\vec{A}_1 \right)^TR_1\label{eq:6.3c}\\
R_2 ^{T}\ddot{R}_2  I_2  -I_2  \ddot{R}_2  ^TR_2  &=&-R^T_2  
\left(
\vec\lambda\otimes\vec{A}_2 
\right)+
\left(
\vec\lambda\otimes\vec{A}_2  \right)^TR_2 \label{eq:6.3d}
\end{eqnarray}
\end{subequations}
The right-hand sides of
equations (\ref{eq:6.3c}, \ref{eq:6.3d}) can be expressed in terms of $\angbod$ in the usual manner
\begin{equation}
\hat{\Theta}_{I_\sigma}(\dot{\Omega}_{b,\sigma})+
\left[
\Omega_{b,\sigma},\hat{\Theta}_{I_\sigma}
(\Omega_{b,\sigma})\right]
\label{eq:6.4}
\end{equation}
Introducing relative and center of mass variables, and taking the
center of mass to be at rest, which is possible due to Galilean invariance, we have
\begin{subequations}
\label{eq:6.5}
\begin{eqnarray}
0&=&M_1\vec{X}_1+M_2\vec{X}_2\label{eq:6.4a}\\
\vec x&=&\vec{X}_1-\vec{X}_2.\label{eq:6.4b}
\end{eqnarray}
\end{subequations}
We may now eliminate $\vec\lambda$ using (\ref{eq:6.3a}, \ref{eq:6.3b}, \ref{eq:6.5}), obtaining, 
after some straightforward algebra
\begin{subequations}
\label{eq:6.6}
\begin{eqnarray}
\hat{\Theta}_{I_1 }(\dot{\Omega}_{b,1 })+
\left[
\Omega_{b,1},\hat{\Theta}_{I_1}
(\Omega_{b,1})\right]
&=&\mu\left[
R^T_1 
\left(\ddot{\vec x}\otimes\vec{A}_1 
\right)-
\left(
\ddot{\vec x}\otimes\vec{A}_1 \right)^TR_1
\right]
\label{eq:6.6a}\\
\hat{\Theta}_{I_2}(\dot{\Omega}_{b,2})+
\left[
\Omega_{b,2},\hat{\Theta}_{I_2}
(\Omega_{b,2})\right]
&=&\mu\left[
\left(
\ddot{\vec x}\otimes\vec{A}_2  \right)^TR_2
-R^T_2  
\left(
\ddot{\vec x}\otimes\vec{A}_2 
\right)
\right]
\label{eq:6.6b} \\
\vec x&=&R_2(t)\vec{A}_2-R_1(t)\vec{A}_1\label{eq:6.6c}\\
\mu&=&\frac{M_1M_2}{M_1+M_2}\label{eq:6.6d}
\end{eqnarray}
\end{subequations}
After substituting (\ref{eq:6.6c}) into (\ref{eq:6.6a}, \ref{eq:6.6b}) one gets---for the three-dimensional 
case---a set of six equations for the 
twelve unknowns $R_\sigma$ and $\Omega_{b,\sigma}$. Together with the equations that define
$\Omega_{b,\sigma}$ in terms of $\dot{R}_ \sigma$ and $R_\sigma$, given by (\ref{eq:2.7a}), one 
obtains a closed set of equations. These remarks extend trivially to the general $n$-dimensional
case.

The physical meaning of these equations is clear: the left-hand sides of Eqs.~(\ref{eq:6.6a}, \ref{eq:6.6b})
are the same as that of the Euler equation with torque, see Eq.~(\ref{eq:4.5}). Their right-hand sides,
on the other hand, express the torque which act on the body $\sigma$ due to the action of the joint, caused by
the relative acceleration between both bodies. 
\section{Conclusions}
\label{sec:conclusion}
In this paper I have primarily focused on free rigid body motion. We have seen how to derive
both a simple equation of motion, namely Eq.~(\ref{eq:3.11}), valid quite generally, as well as Euler's
equation for a free top and the relation between the angular velocity matrix and the conserved angular momentum
via the (generalized) moment of inertia tensor. This way of obtaining
an equation of motion for a system involving one or many rigid bodies is quite general and flexible,
as we have seen in the example of Section \ref{sec:art}.
Extensions to other groups than rotations are also possible, 
as well as to the description of systems such as approximately rigid bodies, for which one introduces
coordinates involving a translation, a rotation and deviations from the reference positions. In all these cases, 
computations quite similar to those described above straightforwardly yield
an equation of motion.

A significant issue with the method developed so far is the absence of a canonical formalism. 
This means that we cannot say which conservation laws are in involution and which are not. 
This severely limits our ability to identify integrable systems. A Hamiltonian formalism 
can, in fact, be developed, but it is by no means as elementary as the Lagrangian formalism 
presented here. Such developments are reserved for a future publication. 

Once the equation of motion has been obtained, we may still proceed to study its conservation laws
without further algebraic difficulties, as we have shown by the application of Noether's theorem to 
the free and the symmetrical top. In fact, we see that the technique described above leads to 
new insights: the fact that the conservation of an appropriate component
of the angular momentum in the body frame follows from the body's  symmetry with respect to the 
corresponding set of rotations, is ordinarily not derived in this fashion.
On the other hand, solving the equation of motion usually requires going to specific coordinates.
This can be arduous, and it may often be simpler to do so directly at the level of the Lagrangian. 
Nevertheless, working on the problem at the abstract level tells us a great deal about its structure,
as I hope to have made clear in the examples presented above. 
Concerning the true usefulness of this approach, however, we might aptly 
 quote one of the fathers of analytical mechanics\cite{hamilton}:
``It may happen to me, as to others, that a meditation which has long been dwelt on shall assume 
an unreal importance; and that a method which has for a long time been practised shall acquire 
an only seeming facility.''

\appendix
\section{Some helpful formulae for calculations with traces}
\label{sec:trace}
The trace of an $n\times n$ matrix is defined as
\begin{equation}
\tr( A)=\sum_{\alpha=1}^nA_{\alpha\alpha}
\label{eq:a1}
\end{equation}
It is readily verified that 
\begin{equation}
\tr (AB)=\sum_{\alpha,\beta=1}^nA_{\alpha\beta}B_{\beta\alpha}=\tr (BA),
\label{eq:a2}
\end{equation}
from which straightforwardly follows
\begin{equation}
\tr (ABC)=\tr (BCA)=\tr( CAB).
\label{eq:a3}
\end{equation}
It follows immediately from (\ref{eq:a2}) that
\begin{equation}
\frac{\partial}{\partial X_{\alpha\beta}}\tr (XY)=Y_{\beta\alpha}
\label{eq:a4}
\end{equation}
which can be symbolically rewritten as
\begin{equation}
\frac{\partial}{\partial X}\tr (XY)=Y^T.
\label{eq:a5}
\end{equation}
Throughout the text we shall often combine (\ref{eq:a3}) and (\ref{eq:a5}) to obtain
such results as
\begin{equation}
\frac{\partial}{\partial Y}\tr
\left(
XYZ^T
\right)
=X^TZ
\label{eq:a6}
\end{equation}

The trace can also be used for other purposes. For example, note that 
\begin{equation}
\tr
\left( 
AB^T
\right)=\sum_{\alpha, \beta}A_{\alpha\beta}B_{\alpha\beta}
\label{eq:a7}
\end{equation}
defines a scalar product on the set of matrices. In fact, it defines the standard
scalar product and we shall often use this fact. 

We also sometimes need to reduce matrix elements of operators to trace form.
This can be done using the concept of tensor product: given two vectors $\vec x$
and $\vec y$, we define $\vec x\otimes\vec y$ as the matrix given by
\begin{equation}
\left(
\vec x\otimes\vec y
\right)_{\alpha\beta}=
x_\alpha y_\beta.
\label{eq:a8}
\end{equation}
From this follows that, for any matrix $A$:
\begin{equation}
\left(
\vec x, A\vec y
\right)=\tr
\left[
\left(
\vec y\otimes\vec x
\right)A
\right]
\label{eq:a9}
\end{equation}

\section{Deriving the usual form of the Euler equations}
\label{sec:euler3d}
It is standard \cite{arnold} that one 
can, to each $3\times3$ antisymmetric matrix $\Omega$, assign 
a vector $\vec\omega\in{\mathbb R}^3$ such that for all $\vec x\in{\mathbb R}^3$
\begin{equation}
\Omega\vec x=\vec\omega\wedge\vec x,
\label{eq:b.1}
\end{equation}
where $\vec x\wedge\vec y$ denotes the usual vector product between $\vec x$ and $\vec y$. 
Using well-known properties of the vector product, we can show the following very
useful equalities:
\begin{subequations}
\label{eq:b.2}
\begin{eqnarray}
\left(
\Omega_1+\Omega_2
\right)\vec x&=&\left(
\vec\omega_1+\vec\omega_2
\right)
\wedge\vec x\label{eq:b.2a}\\
\left(
\Omega_1\Omega_2
-\Omega_2\Omega_1
\right)\vec x&=&\left(
\vec\omega_1\wedge\vec\omega_2
\right)
\wedge\vec x\label{eq:b.2b}\\
R\Omega_1R^{-1}\vec x&=&\left(
R\vec\omega_1
\right)\wedge\vec x\label{eq:b.2c}
\end{eqnarray}
\end{subequations}
for all $\vec x$, where we have assumed
\begin{equation}
\Omega_i\vec x=\vec\omega_i\wedge\vec x
\label{eq:b.3}
\end{equation}
for all $\vec x$. We therefore see that addition and commutation of matrices
translate into addition and vector product of vectors, whereas a change of
coordinates will change the matrix and the vector in compatible ways, 
see Eq.~(\ref{eq:b.2c}). 

The superoperator $\hat{\Theta}_I$ assigns linearly to every 
antisymmetric matrix $\Omega$ the matrix $I\Omega+\Omega I$.
It thus translates into a linear operator $\Theta$ on the vectors
$\vec\omega$. To determine it, start by considering the case 
in which the basis is chosen in such a way as to make $I$ diagonal
(principal axes). In this case, one sees easily
that
\begin{equation}
\left(
\hat{\Theta}_I\Omega
\right)_{\alpha\beta}=
\left(
I_\alpha+I_\beta
\right)\Omega_{\alpha\beta},
\label{eq:b.4}
\end{equation}
where there is {\em no summation\/} over repeated indices. Using 
the explicit form of the transformation of $\Omega$ to $\vec\omega$,
we find
\begin{equation}
\Theta=
\left(
\begin{array}{ccc}
 I_2+I_3 & 0  & 0  \\
0  &  I_1+I_3 &  0 \\
 0 & 0  &  I_1+I_2  
\end{array}
\right)
.
\label{eq:b.5}
\end{equation}
This can be expressed in the form
\begin{equation}
\Theta=\tr\left(
I
 \right)
 \cdot{\mathbb I}
-I.
\label{eq:b.6}
\end{equation}
Since this is an expression which transforms under rotations in the same way as $\Theta$,
namely as a tensor, Eq.~(\ref{eq:b.6}) is generally true. The matrix
$\Theta$ which acts on vectors $\vec\omega$
in the same way as the superoperator $\hat{\Theta}_I$ does on antisymmetric matrices,
is thus given by the usual expression for the moment of inertia tensor. 

All we now need to do is to use Eqs.~(\ref{eq:b.2}) to translate the Euler equations
(\ref{eq:4.5}) derived in Section \ref{sec:euler} into an equation for vectors $\vec\omega$.
One obtains:
\begin{equation}
\Theta\dot{\vec{\omega}}+
\vec\omega
\wedge
\left(
\Theta\vec\omega
\right)
=0.
\label{eq:b.7}
\end{equation}
It goes without saying that the vector $\vec\omega$ refers to the antisymmetric matrix
$\angbod$, that is, to the angular velocity in the body system. 

\begin{acknowledgments}
Helpful discussions with A.~Reyes and A.~Botero are very gratefully acknowledged,
as is the support of DGAPA IN114014 as well as CONACyT 154586.
\end{acknowledgments}


\begin{thebibliography}{99}
\bibitem{foot} One might even consider  integrating (\ref{eq:3.11}) numerically in Cartesian 
coordinates, that is, using all nine entries of the matrix $R(t)$ as coordinates and trusting 
to the nature of the equation to keep $R(t)$ a rotation. This is in  principle possible,
but numerical instabilities might be a problem.
\end{thebibliography}
\end{document}